\documentstyle[prd,aps,preprint,tighten,epsfig]{revtex}

\begin{document}

\draft

\title{Isomeric Lepton Mass Matrices and Bi-large Neutrino Mixing}
\author{\bf Zhi-zhong Xing}
\address{Institute of High Energy Physics, Chinese Academy of Sciences, \\
P.O. Box 918 (4), Beijing 100039, China
\footnote{Mailing address} \\
({\it Electronic address: xingzz@mail.ihep.ac.cn}) }
\author{\bf Shun Zhou}
\address{Department of Physics, Nankai University, Tianjin 300071, China \\
({\it Electronic address: zs00208@phys.nankai.edu.cn}) }
\maketitle

\begin{abstract}
We show that there exist six parallel textures of the charged
lepton and neutrino mass matrices with six vanishing entries, whose
phenomenological consequences are exactly the same. These
{\it isomeric} lepton mass matrices are compatible with current
experimental data at the $3\sigma$ level. If the seesaw mechanism
and the Fukugita-Tanimoto-Yanagida hypothesis are taken into
account, it will be possible to fit the experimental data at or
below the $2\sigma$ level. In particular, the maximal atmospheric neutrino
mixing can be reconciled with a strong neutrino mass hierarchy in
the seesaw case.
\end{abstract}


The recent solar \cite{SNO}, atmospheric \cite{SK}, KamLAND \cite{KM}
and K2K \cite{K2K} neutrino oscillation experiments have provided us with
very convincing evidence that neutrinos are massive and lepton flavors
are mixed. In particular, the admixture of three lepton flavors involves
two large angles $\theta_{12} \sim 33^\circ$ and
$\theta_{23} \sim 45^\circ$ \cite{FIT}. To interpret the observed
{\it bi-large} lepton flavor mixing pattern, many phenomenological
ans$\rm\ddot{a}$tze of lepton mass matrices have been proposed in
the literature \cite{Review}.
A very interesting category of the ans$\rm\ddot{a}$tze focus on
{\it texture zeros} of charged lepton and neutrino mass matrices in a
specific flavor basis, from which some nontrivial and testable
relations between
flavor mixing angles and lepton mass ratios can be derived. A typical
example is the Fritzsch ansatz \cite{F78} of lepton mass matrices,
\begin{equation}
M^{~}_{l,\nu} \; = \; \left ( \matrix{
{\bf 0} & \times    & {\bf 0} \cr
\times  & {\bf 0}   & \times \cr
{\bf 0} & \times    & \times \cr} \right ) \; ,
%   (1)
\end{equation}
in which six texture zeros are included
%%%%%%%%%%%%%%%%%%%%%%%%%%%%%%%%%%%%%%%%
\footnote{Because $M_l$ and $M_\nu$ are taken to be symmetric,
a pair of off-diagonal texture zeros in $M_l$ or $M_\nu$ have
been counted as one zero.}
%%%%%%%%%%%%%%%%%%%%%%%%%%%%%%%%%%%%%%%
and all non-vanishing entries are simply symbolized by $\times$'s.
It has been shown in Ref. \cite{Xing02} that this ansatz can
naturally predict a normal but weak neutrino mass hierarchy and a
bi-large lepton flavor mixing pattern. If the seesaw mechanism is
incorporated in the Fritzsch texture of charged lepton and Dirac
neutrino mass matrices \cite{FTY}, one may obtain a similar flavor
mixing pattern together with a much stronger neutrino mass hierarchy.

The simplicity and predictability of $M_l$ and $M_\nu$ in Eq. (1)
motivate us to examine other possible six-zero textures of lepton
mass matrices and their various phenomenological consequences.
We find that there totally exist six {\it parallel} patterns of
$M_l$ and $M_\nu$ with six texture zeros, as listed in Table 1,
where the Fritzsch ansatz is labelled as pattern (A). It is apparent
that these six patterns are structurally different from one another.
The question is whether their predictions for neutrino masses,
flavor mixing angles and CP violation are distinguishable or not.

The purpose of this paper is to answer the above question and to
confront those six-zero textures of lepton mass matrices with the
latest experimental data. First, we shall present a concise
analysis of the lepton mass matrices in Table 1 and reveal their
{\it isomeric} features -- namely, they have the same phenomenological
consequences, although their structures are apparently different.
Second, we shall examine the predictions of these lepton mass
matrices by comparing them with the $2\sigma$ and $3\sigma$ intervals
of two neutrino mass-squared differences and three lepton flavor
mixing angles \cite{FIT2}, which are obtained from a global analysis of
the latest solar, atmospheric, reactor (KamLAND and CHOOZ \cite{CHOOZ})
and accelerator (K2K) neutrino data. We find no parameter space allowed
for six isomeric lepton mass matrices at the $2\sigma$ level. At the
$3\sigma$ level, however, their results for neutrino masses and
lepton flavor mixing angles can be compatible with current data.
Third, we incorporate the seesaw mechanism and the
Fukugita-Tanimoto-Yanagida hypothesis \cite{FTY} in the charged lepton
and Dirac neutrino mass matrices with six texture zeros. It turns out
that their predictions, including $\theta_{23} \approx 45^\circ$,
are in good agreement with the present experimental data even at the
$2\sigma$ level.

Let us begin with the diagonalization of $M_l$ and $M_\nu$ listed in
Table 1. Without loss of generality, one may take their diagonal
non-vanishing elements to be real and positive. Then only the
off-diagonal non-vanishing elements of $M_l$ and $M_\nu$ are complex.
Each mass matrix $M$ consists of two phase parameters
($\phi$ and $\varphi$) and three real and positive parameters
($A$, $B$ and $C$), as shown in Table 1, where their subscript ``$l$''
or ``$\nu$'' has been omitted for simplicity. The diagonalization of
$M$ requires the following unitary transformation,
\begin{equation}
U^\dagger M U^* \; = \; \left ( \matrix{
\lambda_1 & 0 & 0 \cr
0 & \lambda_2 & 0 \cr
0 & 0 & \lambda_3 \cr} \right ) \; ,
%   (2)
\end{equation}
where $\lambda_i$ (for $i=1,2,3$) denote the physical masses
of charged leptons (i.e., $\lambda_{1,2,3} = m_{e,\mu,\tau}$)
or neutrinos (i.e., $\lambda_i = m_i$). Due to the particular
texture of $M$, $U$ can be written as a product of a diagonal phase
matrix (dependent on $\phi$ and $\varphi$) and a unitary matrix
(independent of $\phi$ and $\varphi$), as illustrated by Table 1.
The real parameters $(A,B,C)$ in $M$ and
$(a^{~}_i,b^{~}_i,c^{~}_i)$ in $U$ are simple functions of $\lambda_i$:
\begin{eqnarray}
A & = & \lambda_3 \left ( 1 - y + xy \right ) \; ,
\nonumber \\
B & = & \lambda_3 \left [ \frac{y (1 - x) (1 - y) (1 + xy)}
{1 - y + xy} \right ]^{1/2} \; ,
\nonumber \\
C & = & \lambda_3 \left ( \frac{x y^2}{1 - y + xy} \right )^{1/2} \; ;
%   (3)
\end{eqnarray}
and
\begin{eqnarray}
a^{~}_1 & = & + \left [ \frac{1-y}{(1+x)(1-xy)(1-y+xy)} \right ]^{1/2} \; ,
\nonumber \\
a^{~}_2 & = & -i \left [ \frac{x(1+xy)}{(1+x)(1+y)(1-y+xy)} \right ]^{1/2} \; ,
\nonumber \\
a^{~}_3 & = & + \left [ \frac{xy^3 (1-x)}{(1-xy)(1+y)(1-y+xy)}
\right ]^{1/2} \; ,
\nonumber \\
b^{~}_1 & = & + \left [ \frac{x(1-y)}{(1+x)(1-xy)} \right ]^{1/2} \; ,
\nonumber \\
b^{~}_2 & = & +i \left [ \frac{1+xy}{(1+x)(1+y)} \right ]^{1/2} \; ,
\nonumber \\
b^{~}_3 & = & + \left [ \frac{y(1-x)}{(1-xy)(1+y)} \right ]^{1/2} \; ,
\nonumber \\
c^{~}_1 & = & - \left [ \frac{xy(1-x)(1+xy)}{(1+x)(1-xy)(1-y+xy)}
\right ]^{1/2} \; ,
\nonumber \\
c^{~}_2 & = & -i \left [ \frac{y(1-x)(1-y)}{(1+x)(1+y)(1-y+xy)}
\right ]^{1/2} \; ,
\nonumber \\
c^{~}_3 & = & + \left [ \frac{(1-y)(1+xy)}{(1-xy)(1+y)(1-y+xy)}
\right ]^{1/2} \; ,
%   (4)
\end{eqnarray}
where $x\equiv \lambda_1/\lambda_2$ and $y \equiv \lambda_2/\lambda_3$
have been defined. Note that $a^{~}_2$, $b^{~}_2$ and $c^{~}_2$ are
imaginary, and their nontrivial phases arise from a minus sign of the
determinant of $M$(i.e., ${\rm Det}(M) = - AC^2 e^{2i\varphi}$).
Since the charged lepton masses have precisely been
measured \cite{PDG}, we have $x^{~}_l \approx 0.00484$ and
$y^{~}_l \approx 0.0594$. On the other hand, $0 < x_\nu < 1$ is
required by the solar neutrino oscillation data \cite{SNO}. Hence
$0 < y_\nu < 1$ must hold, in agreement with Eq. (4). This observation
implies that the isomeric lepton mass matrices under discussion guarantee
a normal neutrino mass spectrum.

The lepton flavor mixing matrix $V$, which links the neutrino mass
eigenstates $(\nu_1, \nu_2, \nu_3)$ to the neutrino flavor eigenstates
$(\nu_e, \nu_\mu, \nu_\tau)$, results from
the mismatch between the diagonalization of $M_l$ and that of
$M_\nu$. Taking account of Eq. (2), we obtain
$V = U^\dagger_l U_\nu$, whose nine matrix
elements read explicitly as
\begin{equation}
V_{pq} \; = \; (a^{l}_p)^* a^\nu_q e^{i\alpha} +
(b^{l}_p)^* b^\nu_q e^{i \beta} + (c^{l}_p)^* c^\nu_q \; ,
%   (5)
\end{equation}
where the subscripts $p$ and $q$ run respectively over $(e, \mu, \tau)$
and $(1,2,3)$, and the phase parameters $\alpha$ and $\beta$ are defined
by $\alpha \equiv (\varphi^{~}_\nu - \varphi^{~}_l) - \beta$ and
$\beta \equiv (\phi^{~}_\nu - \phi^{~}_l)$. It is worth remarking that
Eq. (5) is universally valid for all six patterns of lepton mass matrices
in Table 1. Hence they must have the same phenomenological consequences
and can be referred to as the {\it isomeric} lepton mass matrices.

Obviously, $V$ consists of four unknown parameters:
$x_\nu$, $y_\nu$, $\alpha$ and $\beta$. Their magnitudes can be
constrained by current experimental data on neutrino oscillations.
For the sake of convenience, we adopt the standard parametrization
of $V$ \cite{X04}:
\begin{equation}
V \; = \; \left ( \matrix{
c_{12} c_{13} & s_{12} c_{13} & s_{13} \cr
- c_{12} s_{23} s_{13} - s_{12} c_{23} e^{-i\delta} &
- s_{12} s_{23} s_{13} + c_{12} c_{23} e^{-i\delta} &
s_{23} c_{13} \cr
- c_{12} c_{23} s_{13} + s_{12} s_{23} e^{-i\delta} &
- s_{12} c_{23} s_{13} - c_{12} s_{23} e^{-i\delta} &
c_{23} c_{13} \cr } \right )
\left ( \matrix{
e^{i\rho}   & 0 & 0 \cr
0   & e^{i\sigma}   & 0 \cr
0   & 0 & 1 \cr} \right ) \; ,
%       (6)
\end{equation}
where $c_{ij} \equiv \cos\theta_{ij}$ and $s_{ij} \equiv \sin\theta_{ij}$
(for $ij=12,23,13$). Table 2 is a summary of the allowed ranges of
two neutrino mass-squared differences ($\Delta m^2_{21} \equiv m^2_2 - m^2_1$
and $\Delta m^2_{31} \equiv m^2_3 - m^2_1$) and three flavor mixing angles
($\sin^2 \theta_{12}$, $\sin^2 \theta_{23}$ and $\sin^2 \theta_{13}$),
obtained from a gobal analysis of the latest solar, atmospheric, reactor
and accelerator neutrino data \cite{FIT2}. Because
\begin{equation}
R_\nu \; \equiv \; \frac{\Delta m^2_{21}}{\Delta m^2_{31}}
\; = \; y^2_\nu ~ \frac{1 - x^2_\nu}{1 - x^2_\nu y^2_\nu} \;
%   (7)
\end{equation}
and
\begin{eqnarray}
\sin^2\theta_{12} & = & \frac{|V_{e2}|^2}{1 - |V_{e3}|^2} \; ,
\nonumber \\
\sin^2\theta_{23} & = & \frac{|V_{\mu 3}|^2}{1 - |V_{e3}|^2} \; ,
\nonumber \\
\sin^2\theta_{13} & = & |V_{e3}|^2
%       (8)
\end{eqnarray}
are all dependent on $x_\nu$, $y_\nu$, $\alpha$ and $\beta$,
the latter can then be constrained by using the experimental
data in Table 2. Once the parameter space of $(x_\nu, y_\nu)$ and
$(\alpha, \beta)$ is fixed, one may quantitatively determine the
CP-violating phases $(\delta, \rho, \sigma)$ and the Jarlskog
invariant $\cal J$ ($= {\rm Im} [V_{e2} V_{\mu 3} V^*_{e3} V^*_{\mu 2}]$,
for example \cite{J}), which measures the strength of CP and T
violation in neutrino oscillations. It is also possible to determine
the neutrino mass spectrum and the effective masses of the tritium
beta decay ($\langle m\rangle_e \equiv m_1 |V_{e1}|^2 +
m_2 |V_{e2}|^2 + m_3 |V_{e3}|^2$) and the neutrinoless double beta decay
($\langle m\rangle_{ee} \equiv |m_1 V^2_{e1} + m_2 V^2_{e2} +
m_3 V^2_{e3}|$). The results of our numerical calculations are
summarized as follows.

(1) We find that the parameter space of $(x_\nu, y_\nu)$ or
$(\alpha, \beta)$ will be empty, if the best-fit values or the $2\sigma$
intervals of $\Delta m^2_{21}$, $\Delta m^2_{31}$,
$\sin^2 \theta_{12}$, $\sin^2 \theta_{23}$ and $\sin^2 \theta_{13}$
are taken into account. This situation is caused by the conflict
between the largeness of $\sin^2\theta_{23}$ and the smallness of
$R_\nu$, which cannot simultaneously be achieved from $M_l$ and
$M_\nu$ at the $2\sigma$ level.

(2) If the $3\sigma$ intervals of $\Delta m^2_{21}$, $\Delta m^2_{31}$,
$\sin^2 \theta_{12}$, $\sin^2 \theta_{23}$ and $\sin^2 \theta_{13}$
are taken into account, however, the consequences of $M_l$ and
$M_\nu$ on neutrino masses and flavor mixing angles can be compatible
with current experimental data. Fig. 1 shows the allowed parameter
space of $(x_\nu, y_\nu)$ and $(\alpha, \beta)$ at the $3\sigma$ level.
We see that $\beta \sim \pi$ holds. This result is consistent with the
previous observation \cite{Xing02}. Because of $y_\nu \sim 0.25$,
$m_3 \approx \sqrt{\Delta m^2_{31}}$ is a good approximation. The
neutrino mass spectrum can actually be determined to an acceptable
degree of accuracy:
$m_3 \approx (3.8 - 6.1) \times 10^{-2}$ eV,
$m_2 \approx (0.95 - 1.5) \times 10^{-2}$ eV and
$m_1 \approx (2.6 - 3.4) \times 10^{-3}$ eV, where $x_\nu \approx 1/3$
and $y_\nu \approx 1/4$ have typically be taken.
A straightforward calculation yields
$\langle m\rangle_e \sim 10^{-2} ~ {\rm eV}$ for the tritium beta
decay and $\langle m\rangle_{ee} \sim 10^{-3} ~ {\rm eV}$ for the
neutrinoless double beta decay. Both of them are too small to be
experimentally accessible in the foreseeable future.

(3) Fig. 2 shows the outputs of $\sin^2 \theta_{12}$,
$\sin^2 \theta_{23}$ and $\sin^2 \theta_{13}$ versus $R_\nu$ at the
$3\sigma$ level. It is obvious that the maximal atmospheric neutrino
mixing (i.e., $\sin^2\theta_{23} \approx 0.5$ or
$\sin^2 2\theta_{23} \approx 1$) cannot be achieved from
the isomeric lepton mass matrices under consideration. We see that
$\sin^2\theta_{23} < 0.40$ (or $\sin^2 2\theta_{23} < 0.96$)
holds in our ansatz, and it is impossible
to get a larger value of $\sin^2\theta_{23}$ even if $R_\nu$
approaches its upper bound. In contrast, the output of $\sin^2\theta_{12}$
is favorable and has less dependence on $R_\nu$. One can also see that
only small values of $\sin^2\theta_{13}$ ($\leq 0.016$) are favored.
More precise data on $\sin^2\theta_{23}$, $\sin^2\theta_{13}$ and $R_\nu$
will allow us to check whether those isomeric lepton mass
matrices with six texture zeros can really survive the experimental
test or not.

(4) We calculate the CP-violating phases $(\delta, \rho, \sigma)$ and
the Jarlskog invariant $\cal J$, and illustrate their results in Fig. 3.
The maximal magnitude of $\cal J$ is close to 0.015 around
$\delta \sim 3\pi/4$ or $5\pi/4$. As for the Majorana phases $\rho$
and $\sigma$, the relation $(\rho - \sigma) \approx \pi/2$ holds.
This result is attributed to the fact that the matrix elements
$(a^\nu_2, b^\nu_2, c^\nu_2)$ of $U_\nu$ are all imaginary and they
give rise to an irremovable phase shift between $V_{p1}$ and $V_{p2}$
(for $p=e, \mu, \tau$) elements through Eq. (5). Such a phase difference
may affect the effective mass of the neutrinoless double beta decay,
but it has nothing to do with CP violation in neutrino oscillations.

We proceed to discuss a simple way to avoid the potential tension
between the smallness of $R_\nu$ and the largeness of $\sin^2\theta_{23}$
arising from the above isomeric lepton mass matrices. In this connection,
we take account of the Fukugita-Tanimoto-Yanagida hypothesis \cite{FTY}
together with the seesaw mechanism \cite{SS} -- namely,
the charged lepton mass matrix $M_l$ and the Dirac neutrino mass matrix
$M_{\rm D}$ may take one of the six patterns illustrated in Table 1,
while the right-handed Majorana neutrino mass matrix $M_{\rm R}$ takes
the form $M_{\rm R} = M_0 {\bf I}$ with $M_0$ being a very large mass
scale and $\bf I$ denoting the unity matrix.
Then the effective (left-handed) neutrino mass matrix $M_\nu$ reads as
\begin{equation}
M_\nu \; =\; M_{\rm D} M^{-1}_R M^T_{\rm D} \; =\;
\frac{M^2_{\rm D}}{M_0} \; .
%       (9)
\end{equation}
For simplicity, we further assume $M_{\rm D}$ to be real (i.e.,
$\phi^{~}_{\rm D} = \varphi^{~}_{\rm D} =0$). It turns out that the real
orthogonal transformation $U_{\rm D}$, which is defined to diagonalize
$M_{\rm D}$, can simultaneously diagonalize $M_\nu$:
\begin{equation}
U^T_{\rm D} M_\nu U_{\rm D} \; =\;
\frac{(U^T_{\rm D} M_{\rm D} U_{\rm D})^2}{M_0} \; =\;
\left ( \matrix{
m_1 & 0 & 0 \cr
0 & m_2 & 0 \cr
0 & 0 & m_3 \cr} \right ) \; ,
%       (10)
\end{equation}
where $m_i \equiv d^2_i/M_0$ with $d_i$ standing for the eigenvalues
of $M_{\rm D}$. In terms of the neutrino mass ratios
$x_\nu \equiv m_1/m_2 = (d_1/d_2)^2$ and
$y_\nu \equiv m_2/m_3 = (d_2/d_3)^2$, we obtain the
explicit expressions of nine matrix elements of $U_\nu = U_{\rm D}$:
\begin{eqnarray}
a^\nu_1 & = & + \left [ \frac{1-\sqrt{y_\nu}}
{(1+\sqrt{x_\nu})(1-\sqrt{x_\nu y_\nu})
(1-\sqrt{y_\nu}+\sqrt{x_\nu y_\nu})} \right ]^{1/2} \; ,
\nonumber \\
a^\nu_2 & = & - \left [ \frac{\sqrt{x_\nu}(1+\sqrt{x_\nu y_\nu})}
{(1+\sqrt{x_\nu})(1+\sqrt{y_\nu})
(1-\sqrt{y_\nu}+\sqrt{x_\nu y_\nu})} \right ]^{1/2} \; ,
\nonumber \\
a^\nu_3 & = & + \left [ \frac{y_\nu \sqrt{x_\nu y_\nu}
(1-\sqrt{x_\nu})}{(1-\sqrt{x_\nu y_\nu})(1+\sqrt{y_\nu})
(1-\sqrt{y_\nu}+\sqrt{x_\nu y_\nu})} \right ]^{1/2} \; ,
\nonumber \\
b^\nu_1 & = & + \left [ \frac{\sqrt{x_\nu}(1-\sqrt{y_\nu})}
{(1+\sqrt{x_\nu})(1-\sqrt{x_\nu y_\nu})} \right ]^{1/2} \; ,
\nonumber \\
b^\nu_2 & = & + \left [ \frac{1+\sqrt{x_\nu y_\nu}}
{(1+\sqrt{x_\nu})(1+\sqrt{y_\nu})} \right ]^{1/2} \; ,
\nonumber \\
b^\nu_3 & = & + \left [ \frac{\sqrt{y_\nu}(1-\sqrt{x_\nu})}
{(1-\sqrt{x_\nu y_\nu})(1+\sqrt{y_\nu})} \right ]^{1/2} \; ,
\nonumber \\
c^\nu_1 & = & - \left [ \frac{\sqrt{x_\nu y_\nu}(1-\sqrt{x_\nu})
(1+\sqrt{x_\nu y_\nu})}{(1+\sqrt{x_\nu})(1-\sqrt{x_\nu y_\nu})
(1-\sqrt{y_\nu}+\sqrt{x_\nu y_\nu})} \right ]^{1/2} \; ,
\nonumber \\
c^\nu_2 & = & - \left [ \frac{\sqrt{y_\nu}(1-\sqrt{x_\nu})
(1-\sqrt{y_\nu})}{(1+\sqrt{x_\nu})(1+\sqrt{y_\nu})
(1-\sqrt{y_\nu}+\sqrt{x_\nu y_\nu})} \right ]^{1/2} \; ,
\nonumber \\
c^\nu_3 & = & + \left [ \frac{(1-\sqrt{y_\nu})(1+\sqrt{x_\nu y_\nu})}
{(1-\sqrt{x_\nu y_\nu})(1+\sqrt{y_\nu})(1-\sqrt{y_\nu}+\sqrt{x_\nu y_\nu})}
\right ]^{1/2} \; .
%   (11)
\end{eqnarray}
The lepton flavor mixing matrix $V = U^\dagger_l U_\nu$ remains to
take the same form as Eq. (5), but the relevant phase parameters
are now defined as $\alpha \equiv -\varphi^{~}_l -\beta$ and
$\beta \equiv - \phi^{~}_l$.
Comparing between Eqs. (4) and (11), we immediately see that
the magnitudes of $(\theta_{12}, \theta_{23}, \theta_{13})$ in the
non-seesaw case can be reproduced in the seesaw case with much smaller
values of $x_\nu$ and $y_\nu$. The latter will allow $R_\nu$ to be
more strongly suppressed. It is therefore possible to relax the
tension between the smallness of $R_\nu$ and the largeness
of $\sin^2\theta_{23}$ appearing in the non-seesaw case. A careful
numerical analysis of six seesaw-modified patterns of the isomeric lepton
mass matrices {\it does} support this observation. We summarize the
results of our calculations as follows.

(a) We find that the new ansatz are compatible very well with current
neutrino oscillation data, even if the $2\sigma$ intervals of
$\Delta m^2_{21}$, $\Delta m^2_{31}$, $\sin^2 \theta_{12}$,
$\sin^2 \theta_{23}$ and $\sin^2 \theta_{13}$ are taken into account.
Hence it is unnecessary to do a similar analysis at the $3\sigma$
level. The parameter space of $(x_\nu, y_\nu)$ and $(\alpha, \beta)$
is illustrated in Fig. 4, where $x_\nu \sim y_\nu \sim 0.2$ and
$\beta \sim \pi$ hold approximately.
Again $m_3 \approx \sqrt{\Delta m^2_{31}}$ is a good approximation. The
values of three neutrino masses read explicitly as
$m_3 \approx (4.2 - 5.8) \times 10^{-2}$ eV,
$m_2 \approx (0.84 - 1.2) \times 10^{-2}$ eV and
$m_1 \approx (1.6 - 1.9) \times 10^{-3}$ eV, which are obtained by
taking $x_\nu \approx y_\nu \approx 0.2$. It is easy to arrive
at $\langle m\rangle_e \sim 10^{-2} ~ {\rm eV}$ for the tritium beta
decay and $\langle m\rangle_{ee} \sim 10^{-3} ~ {\rm eV}$ for the
neutrinoless double beta decay, thus both of them are too small to be
experimentally accessible in the near future.

(b) The outputs of $\sin^2 \theta_{12}$,
$\sin^2 \theta_{23}$ and $\sin^2 \theta_{13}$ versus $R_\nu$ are shown
in Fig. 5 at the $2\sigma$ level. One can see that the magnitude of
$\sin^2 \theta_{12}$ is essentially unconstrained. Now the maximal
atmospheric neutrino mixing (i.e., $\sin^2\theta_{23} \approx 0.5$ or
$\sin^2 2\theta_{23} \approx 1$) is achievable in the region of
$R_\nu \sim 0.036-0.047$. It is also possible to obtain
$\sin^2\theta_{13} \leq 0.035$, just below the experimental upper
bound \cite{CHOOZ}. If $\sin^2 2\theta_{13} \geq 0.02$ really holds,
the measurement of $\theta_{13}$ should be realizable in a future
reactor neutrino oscillation experiment \cite{T13}.

(c) Fig. 6 illustrates the numerical results of $\delta$, $\rho$,
$\sigma$ and $\cal J$. We see that $|{\cal J}| \sim 0.025$ can be
obtained. Such a size of CP violation is expected to be measured
in the future long-baseline neutrino oscillation experiments.
As for the Majorana phases $\rho$ and $\sigma$, the relation
$\sigma \approx \rho$ holds. This result is easily understandable,
because $U_\nu$ is real in the seesaw case. It is worth mentioning
that the effective neutrino mass matrix $M_\nu$ does not persist in
the simple texture as $M_l$ has, thus the allowed ranges of $\delta$,
$\rho$ and $\sigma$ become smaller in the seesaw case than in
the non-seesaw case.

Note that the eigenvalues of $M_{\rm D}$ and the heavy Majorana
mass scale $M_0$ are not specified in the above analysis. But
one may obtain $|d_1/d_2| = \sqrt{x_\nu} \sim 0.4$ and
$|d_2/d_3| = \sqrt{y_\nu} \sim 0.4$. Such a weak hierarchy of
$(|d_1|, |d_2|, |d_3|)$ means that $M_{\rm D}$ cannot directly
be connected to the charged lepton mass matrix $M_l$, nor can it
be related to the up-type quark mass matrix ($M_{\rm u}$) or
its down-type counterpart ($M_{\rm d}$) in a simple way. If the
hypothesis $M_{\rm R} = M_0 {\bf I}$ is rejected but the result
$U^T_\nu M_\nu U_\nu = {\rm Diag}\{m_1, m_2, m_3\}$ with
$U_\nu$ given by Eq. (11) is maintained, it will be possible to
determine the pattern of $M_{\rm R}$ by means of the inverted
seesaw formula
$M_{\rm R} = M^T_{\rm D} M^{-1}_\nu M_{\rm D}$ \cite{XZ} and by
assuming a specific relation between $M_{\rm D}$ and $M_{\rm u}$.
For example, one may simply assume $M_{\rm D} = M_{\rm u}$ with
$M_{\rm u}$ taking the approximate Fritzsch form,
\begin{equation}
M_{\rm u} \; \sim \; \left ( \matrix{
{\bf 0} & \sqrt{m_u m_c} & {\bf 0} \cr
\sqrt{m_u m_c} & {\bf 0} & \sqrt{m_c m_t} \cr
{\bf 0} & \sqrt{m_c m_t} & m_t \cr} \right ) \; .
%       (12)
\end{equation}
Just for the purpose of illustration, we typically input
$x_\nu \sim y_\nu \sim 0.18$ as well as
$m_u/m_c \sim m_c/m_t \sim 0.0031$ and $m_t \approx 175$ GeV at
the electroweak scale \cite{Xing03}. Then we arrive at
\begin{equation}
M_{\rm R} \; \sim \; 3.0 \times 10^{15} \times \left ( \matrix{
6.1 \times 10^{-8} & 1.2 \times 10^{-5} & 2.0 \times 10^{-4} \cr
1.2 \times 10^{-5} & 3.5 \times 10^{-3} & 5.9 \times 10^{-2} \cr
2.0 \times 10^{-4} & 5.9 \times 10^{-2} & {\bf 1} \cr} \right ) \;
%       (13)
\end{equation}
in unit of GeV. This order-of-magnitude estimate shows that the
scale of $M_{\rm R}$ is close to that of grand unified theories
$\Lambda_{\rm GUT} \sim 10^{16}$ GeV, but the texture of $M_{\rm R}$
and that of $M_{\rm D}$ (or $M_l$) have little similarity. It is
certainly a very nontrivial task to combine the seesaw mechanism
and those phenomenologically-favored patterns of lepton mass
matrices. In this sense, the simple scenarios discussed in
Ref. \cite{FTY} and in the present paper may serve as a helpful
example to give readers a ball-park feeling of the problem itself
and possible solutions to it.

In summary, we have analyzed six parallel patterns of lepton mass
matrices with six texture zeros and demonstrated that their
phenomenological consequences are exactly the same. Confronting
the predictions of these isomeric lepton mass matrices with current
neutrino oscillation data, we find that there is no parameter
space at the $2\sigma$ level. They can be compatible with the
experimental data at the $3\sigma$ level, but it is impossible
to obtain the maximal atmospheric neutrino mixing. We have also
discussed a very simple way to incorporate the seesaw mechanism in
the charged lepton and Dirac neutrino mass matrices with six
texture zeros. It is found that there is no problem to fit current
data even at the $2\sigma$ level in the seesaw case. In particular,
the maximal atmospheric neutrino mixing can naturally be reconciled
with a relatively strong neutrino mass hierarchy. The results for the
effective masses of the tritium beta decay and the neutrinoless double
beta decay are too small to be experimentally accessible in both
the seesaw and non-seesaw cases, but the strength of CP violation
can reach the percent level and may be detectable in the future
long-baseline neutrino oscillation experiments.

We conclude that the peculiar feature of isomeric lepton mass matrices
is very suggestive for model building. We therefore look forward to
seeing whether such simple phenomenological ans$\rm\ddot{a}$tze can
survive the more stringent experimental test or not.

\vspace{0.5cm}

One of us (S.Z.) is grateful to the theory division of IHEP for
financial support and hospitality in Beijing. This work was supported
in part by the National Natural Science Foundation of China.

\newpage

%%%%%%%%%%%%%%%%%%%%% Table 1 %%%%%%%%%%%%%%%%%%%%%%%
\begin{table}
\caption{The isomeric lepton mass matrices ($M_l$ and $M_\nu$)
with six texture zeros and the unitary matrices ($U_l$ and
$U_\nu$) used to diagonalize them, where the subscripts
``$l$'' and ``$\nu$'' have been omitted for simplicity.}
\vspace{0.3cm}
\begin{center}
\begin{tabular}{ccc}
%----------------------------------------------------------------------
(A) &
$M = \left ( \matrix{
{\bf 0} & Ce^{i\varphi} & {\bf 0} \cr
Ce^{i\varphi} & {\bf 0} & Be^{i\phi} \cr
{\bf 0} & Be^{i\phi} & A \cr} \right )$ &
$U = \left ( \matrix{
e^{i(\varphi -\phi)} & 0 & 0 \cr
0 & e^{i\phi} & 0 \cr
0 & 0 & 1 \cr} \right )
\left ( \matrix{
a^{~}_1 & a^{~}_2 & a^{~}_3 \cr
b^{~}_1 & b^{~}_2 & b^{~}_3 \cr
c^{~}_1 & c^{~}_2 & c^{~}_3 \cr} \right )$ \\
%--------------------------------------------------
(B) &
$M = \left ( \matrix{
{\bf 0} & {\bf 0} & Ce^{i\varphi} \cr
{\bf 0} & A & Be^{i\phi} \cr
Ce^{i\varphi} & Be^{i\phi} & {\bf 0} \cr} \right )$ &
$U = \left ( \matrix{
e^{i(\varphi -\phi)} & 0 & 0 \cr
0 & 1 & 0 \cr
0 & 0 & e^{i\phi} \cr} \right )
\left ( \matrix{
a^{~}_1 & a^{~}_2 & a^{~}_3 \cr
c^{~}_1 & c^{~}_2 & c^{~}_3 \cr
b^{~}_1 & b^{~}_2 & b^{~}_3 \cr} \right )$ \\
%--------------------------------------------------
(C) &
$M = \left ( \matrix{
{\bf 0} & Ce^{i\varphi} & Be^{i\phi} \cr
Ce^{i\varphi} & {\bf 0} & {\bf 0} \cr
Be^{i\phi} & {\bf 0} & A \cr} \right )$ &
$U = \left ( \matrix{
e^{i\phi} & 0 & 0 \cr
0 & e^{i(\varphi-\phi)} & 0 \cr
0 & 0 & 1 \cr} \right )
\left ( \matrix{
b^{~}_1 & b^{~}_2 & b^{~}_3 \cr
a^{~}_1 & a^{~}_2 & a^{~}_3 \cr
c^{~}_1 & c^{~}_2 & c^{~}_3 \cr} \right )$ \\
%--------------------------------------------------
(D) &
$M = \left ( \matrix{
{\bf 0} & Be^{i\phi} & Ce^{i\varphi} \cr
Be^{i\phi} & A & {\bf 0} \cr
Ce^{i\varphi} & {\bf 0} & {\bf 0} \cr} \right )$ &
$U = \left ( \matrix{
e^{i\phi} & 0 & 0 \cr
0 & 1 & 0 \cr
0 & 0 & e^{i(\varphi-\phi)} \cr} \right )
\left ( \matrix{
b^{~}_1 & b^{~}_2 & b^{~}_3 \cr
c^{~}_1 & c^{~}_2 & c^{~}_3 \cr
a^{~}_1 & a^{~}_2 & a^{~}_3 \cr} \right )$ \\
%--------------------------------------------------
(E) &
$M = \left ( \matrix{
A & {\bf 0} & Be^{i\phi} \cr
{\bf 0} & {\bf 0} & Ce^{i\varphi} \cr
Be^{i\phi} & Ce^{i\varphi} & {\bf 0} \cr} \right )$ &
$U = \left ( \matrix{
1 & 0 & 0 \cr
0 & e^{i(\varphi-\phi)} & 0 \cr
0 & 0 & e^{i\phi} \cr} \right )
\left ( \matrix{
c^{~}_1 & c^{~}_2 & c^{~}_3 \cr
a^{~}_1 & a^{~}_2 & a^{~}_3 \cr
b^{~}_1 & b^{~}_2 & b^{~}_3 \cr} \right )$ \\
%--------------------------------------------------
(F) &
$M = \left ( \matrix{
A & Be^{i\phi} & {\bf 0} \cr
Be^{i\phi} & {\bf 0} & Ce^{i\varphi} \cr
{\bf 0} & Ce^{i\varphi} & {\bf 0} \cr} \right )$ &
$U = \left ( \matrix{
1 & 0 & 0 \cr
0 & e^{i\phi} & 0 \cr
0 & 0 & e^{i(\varphi-\phi)} \cr} \right )
\left ( \matrix{
c^{~}_1 & c^{~}_2 & c^{~}_3 \cr
b^{~}_1 & b^{~}_2 & b^{~}_3 \cr
a^{~}_1 & a^{~}_2 & a^{~}_3 \cr} \right )$ \\
%-------------------------------------------------
\end{tabular}
\end{center}
\end{table}
%%%%%%%%%%%%%%%%%%%%%%%%%%%%%%%%%%%%%%%%%%%%%%%%%

%%%%%%%%%%%%%%%%%%%%% Table 2 %%%%%%%%%%%%%%%%%%%%%%%
\begin{table}
\caption{The best-fit values, $2\sigma$ and $3\sigma$ intervals
of $\Delta m^2_{21}$, $\Delta m^2_{31}$, $\sin^2 \theta_{12}$,
$\sin^2 \theta_{23}$ and $\sin^2 \theta_{13}$ obtained from a
global analysis of the latest solar, atmospheric, reactor and
accelerator neutrino oscillation data [10].}
\vspace{0.3cm}
\begin{center}
\begin{tabular}{c|ccccc}
%----------------------------------------------------------------------
& $\Delta m^2_{21}$ ($10^{-5} ~ {\rm eV}^2$)
& $\Delta m^2_{31}$ ($10^{-3} ~ {\rm eV}^2$)
& $\sin^2 \theta_{12}$
& $\sin^2 \theta_{23}$
& $\sin^2 \theta_{13}$ \\ \hline
%-----------------------------------------------------
Best fit & 6.9 & 2.6 & 0.30 & 0.52 & 0.006 \cr
2$\sigma$ & 6.0--8.4 & 1.8--3.3 & 0.25--0.36 & 0.36--0.67 & $\leq$ 0.035 \cr
3$\sigma$ & 5.4--9.5 & 1.4--3.7 & 0.23--0.39 & 0.31--0.72 & $\leq$ 0.054
%-------------------------------------------------
\end{tabular}
\end{center}
\end{table}
%%%%%%%%%%%%%%%%%%%%%%%%%%%%%%%%%%%%%%%%%%%%%%%%%

\newpage

%%%%%%%%%%%%%%%%%%%% Fig. 1 %%%%%%%%%%%%%%%%
\begin{figure}[t]
\vspace{-2cm}
\epsfig{file=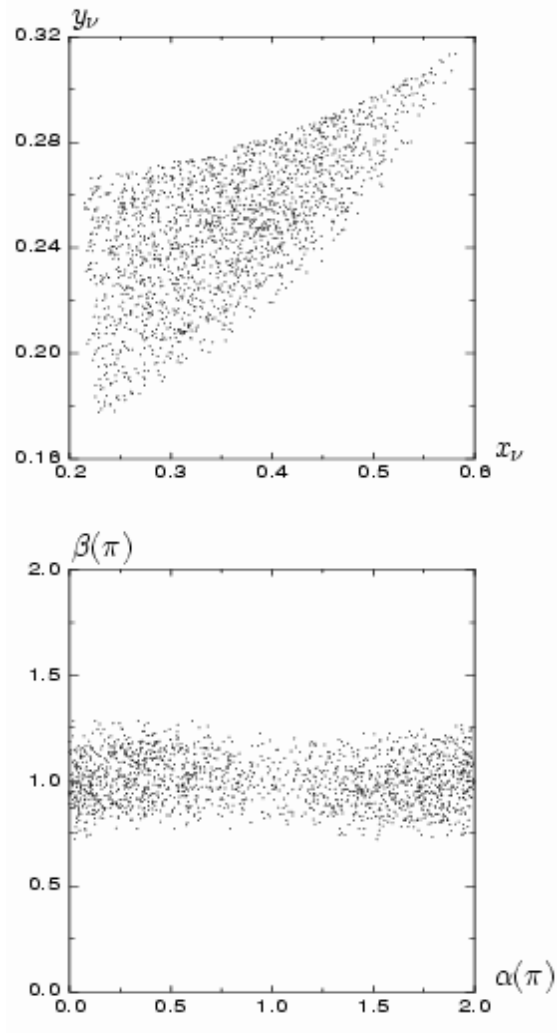,bbllx=-7cm,bblly=14cm,bburx=-2cm,bbury=30cm,%
width=3cm,height=10cm,angle=0,clip=0} \vspace{10cm} \caption{The
parameter space of $(x_\nu, y^{~}_\nu)$ and $(\alpha, \beta)$ at
the $3\sigma$ level.}
\end{figure}
%%%%%%%%%%%%%%%%%%%%%%%%%%%%%%%%%%%%%%%%%%%

%%%%%%%%%%%%%%%%%%%% Fig. 2 %%%%%%%%%%%%%%%%
\begin{figure}[t]
\vspace{-2cm}
\epsfig{file=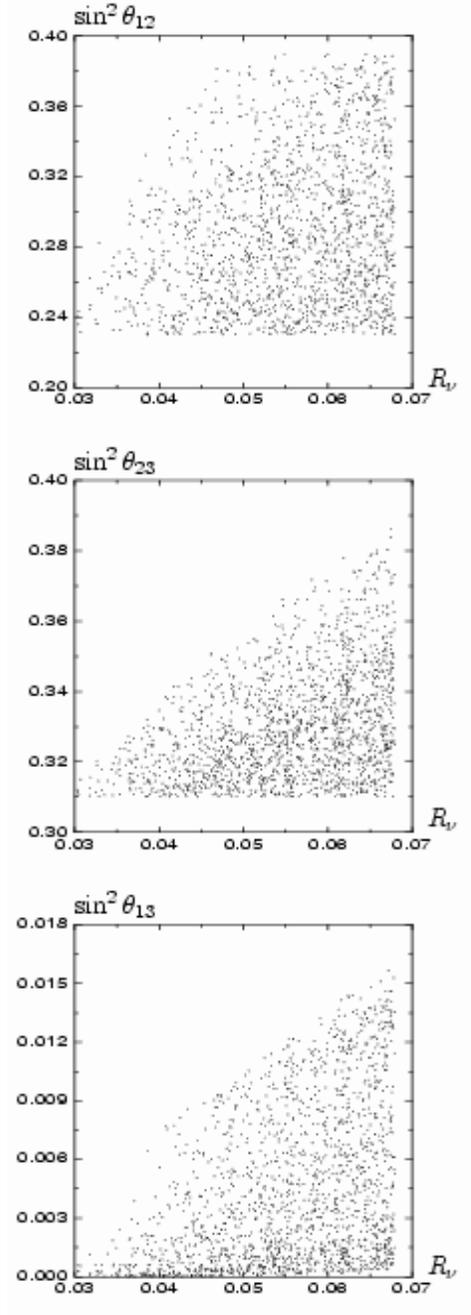,bbllx=-8cm,bblly=14cm,bburx=-3cm,bbury=30cm,%
width=3cm,height=10cm,angle=0,clip=0} \vspace{10cm} \caption{The
outputs of $\sin^2 \theta_{12}$, $\sin^2 \theta_{23}$ and $\sin^2
\theta_{13}$ versus $R_\nu$ at the $3\sigma$ level.}
\end{figure}
%%%%%%%%%%%%%%%%%%%%%%%%%%%%%%%%%%%%%%%%%%%

%%%%%%%%%%%%%%%%%%%% Fig. 3 %%%%%%%%%%%%%%%%
\begin{figure}[t]
\vspace{-2cm}
\epsfig{file=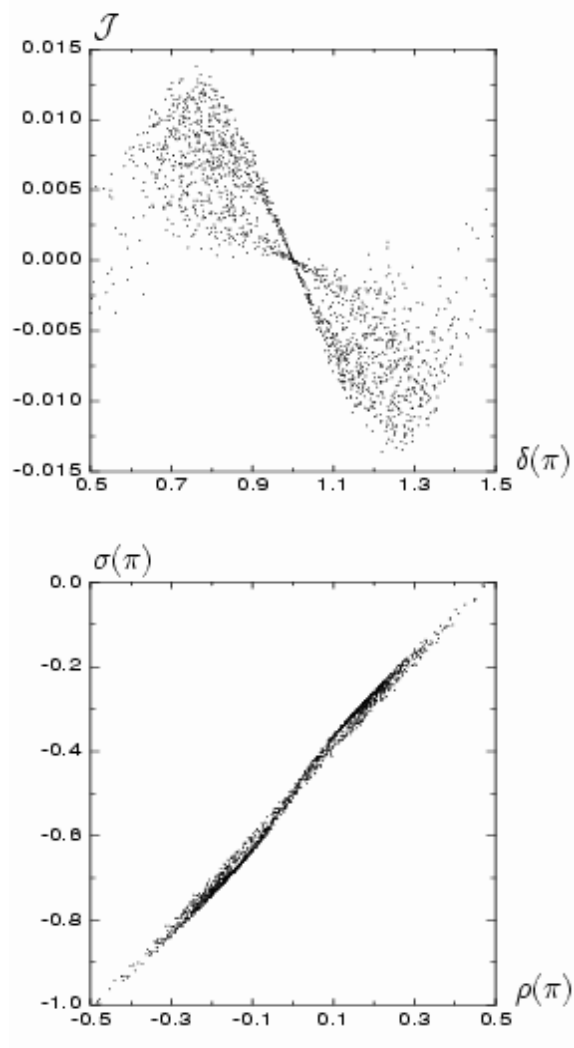,bbllx=-7cm,bblly=14cm,bburx=-2cm,bbury=30cm,%
width=3cm,height=10cm,angle=0,clip=0} \vspace{10cm} \caption{The
outputs of $(\delta, {\cal J})$ and $(\rho, \sigma)$ at the
$3\sigma$ level.}
\end{figure}
%%%%%%%%%%%%%%%%%%%%%%%%%%%%%%%%%%%%%%%%%%%

%%%%%%%%%%%%%%%%%%%% Fig. 4 %%%%%%%%%%%%%%%%
\begin{figure}[t]
\vspace{-2cm}
\epsfig{file=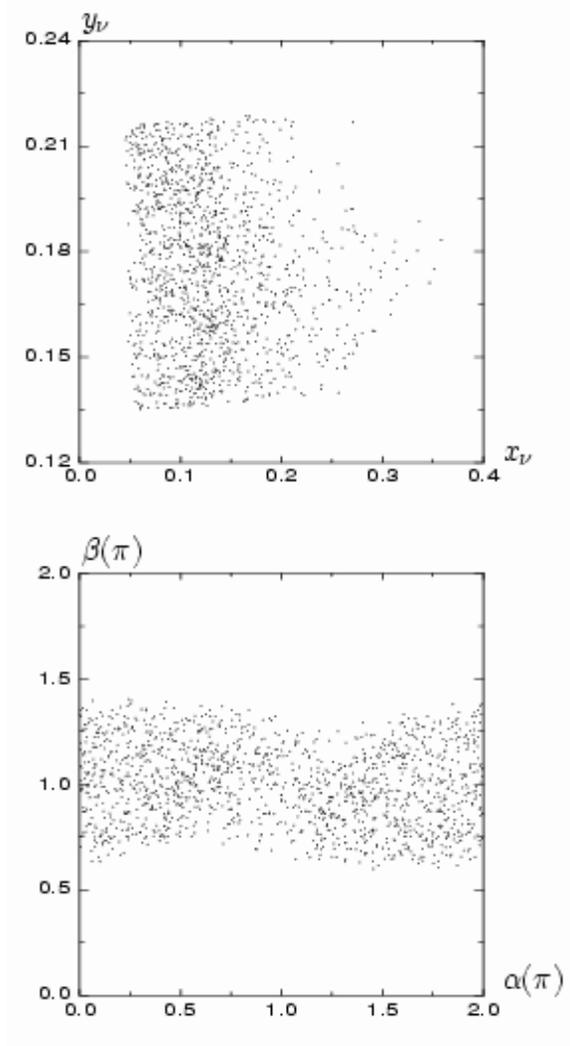,bbllx=-7cm,bblly=14cm,bburx=-2cm,bbury=30cm,%
width=3cm,height=10cm,angle=0,clip=0} \vspace{10cm} \caption{The
parameter space of $(x_\nu, y^{~}_\nu)$ and $(\alpha, \beta)$ at
the $2\sigma$ level in the seesaw case.}
\end{figure}
%%%%%%%%%%%%%%%%%%%%%%%%%%%%%%%%%%%%%%%%%%%

%%%%%%%%%%%%%%%%%%%% Fig. 2 %%%%%%%%%%%%%%%%
\begin{figure}[t]
\vspace{-2cm}
\epsfig{file=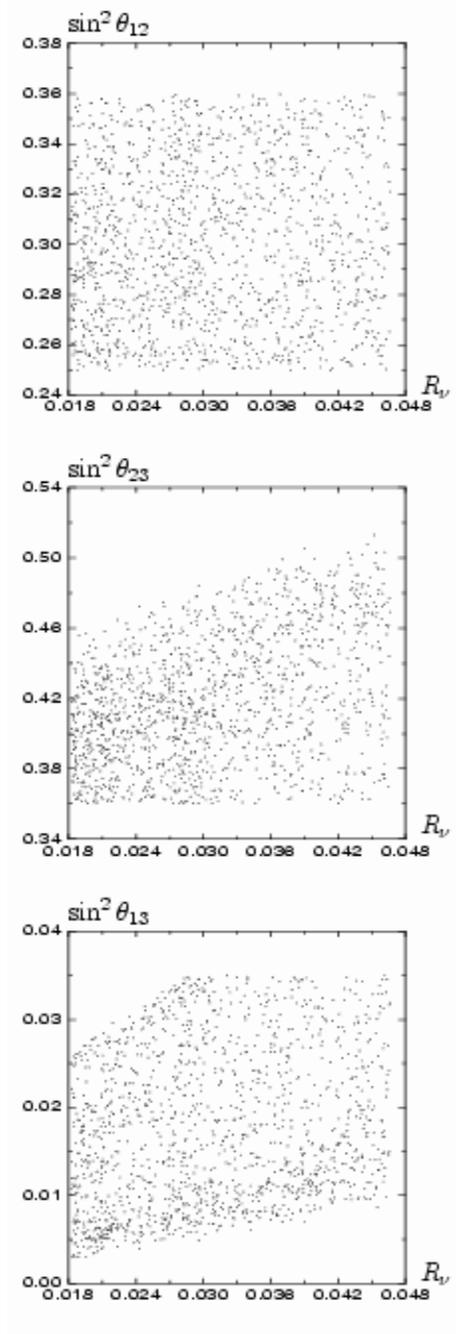,bbllx=-8cm,bblly=14cm,bburx=-3cm,bbury=30cm,%
width=3cm,height=10cm,angle=0,clip=0} \vspace{10cm} \caption{The
outputs of $\sin^2 \theta_{12}$, $\sin^2 \theta_{23}$ and $\sin^2
\theta_{13}$ versus $R_\nu$ at the $2\sigma$ level in the seesaw
case.}
\end{figure}
%%%%%%%%%%%%%%%%%%%%%%%%%%%%%%%%%%%%%%%%%%%

%%%%%%%%%%%%%%%%%%%% Fig. 3 %%%%%%%%%%%%%%%%
\begin{figure}[t]
\vspace{-2cm}
\epsfig{file=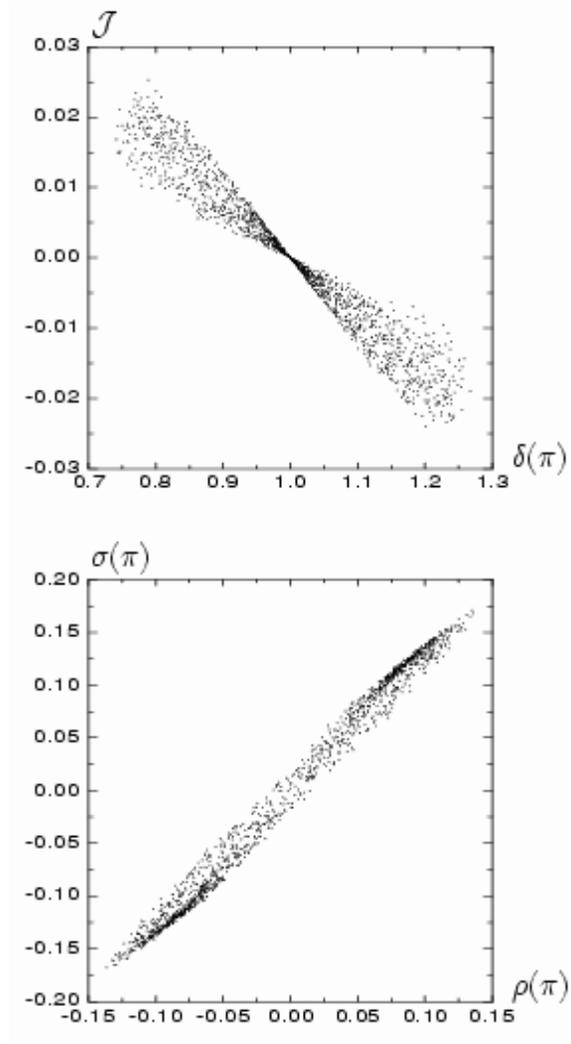,bbllx=-7cm,bblly=14cm,bburx=-2cm,bbury=30cm,%
width=3cm,height=10cm,angle=0,clip=0} \vspace{10cm} \caption{The
outputs of $(\delta, {\cal J})$ and $(\rho, \sigma)$ at the
$2\sigma$ level in the seesaw case.}
\end{figure}
%%%%%%%%%%%%%%%%%%%%%%%%%%%%%%%%%%%%%%%%%%%

\end{document}